\documentclass[12pt,a4paper]{article}

\usepackage{amsmath,amsfonts,latexsym,amssymb}
\usepackage{verbatim,amsthm,curves,graphics}
\usepackage{mathrsfs}

\renewcommand{\d}{{\rm d}}
\newcommand{\EB}{E_{\rm B}}
\newcommand{\MB}{M_{\rm B}}
\newcommand{\PB}{{\bf P}_{\rm B}}

\renewcommand{\v}{{\bf v}}
\newcommand{\x}{{\bf x}}
\newcommand{\z}{{\bf z}}

\newcommand{\sI}{{\mathscr I}}
\newcommand{\ga}{\gamma}
\newcommand{\M}{{\mathscr M}}
\newcommand{\half}{\frac{1}{2}}
\newcommand{\F}{{\mathcal F}}

\def\ben{\begin{equation}}
\def\een{\end{equation}}
\def\bena{\begin{eqnarray}}
\def\eena{\end{eqnarray}}

\newtheorem{lemma}{Lemma}
\newtheorem{proposition}[lemma]{Proposition}
\newtheorem{theorem}[lemma]{Theorem}

\theoremstyle{definition}

\begin{document}
\title{BMS Supertranslations and Memory in Four and Higher Dimensions}

\author{Stefan Hollands$^{1}$\thanks{\tt stefan.hollands@uni-leipzig.de}, 
Akihiro Ishibashi$^{2}$\thanks{\tt akihiro@phys.kindai.ac.jp} 
and Robert M. Wald$^{3}$\thanks{\tt rmwa@midway.uchicago.edu} 
\\ \\
{\it ${}^{1}$Institut f\" ur Theoretische Physik,
Universit\" at Leipzig, }\\
{\it Br\" uderstrasse 16, D-04103 Leipzig, Germany} \\
{\it ${}^{2}$Department of Physics, Kindai University, } \\
{\it Higashi-Osaka, 577-8502, Japan} \\
{\it ${}^{3}$Enrico Fermi Institute and Department of Physics,}\\  
{\it The University of Chicago, }\\ 
{\it 5640 South Ellis Avenue, Chicago, Illinois 60637, USA} \\
}

\date{\today}

\maketitle

\begin{abstract}

We consider the memory effect in even dimensional spacetimes of dimension $d \ge 4$ arising from a burst of gravitational radiation. When $d=4$, the natural frames in the stationary eras before and after the burst differ by the composition of a boost and supertranslation, and this supertranslation characterizes the ``memory effect,'' i.e., the permanent displacement of test particles near infinity produced by the radiation burst. However, we show that when $d > 4$, this supertranslation and the corresponding memory effect vanish. Consequently, when $d >4$, it is natural to impose stronger asymptotic conditions at null infinity that reduce the asymptotic symmetry group to the Poincare group. Conversely, when $d=4$, the asymptotic symmetry group at null infinity must be taken to be the BMS group.

\end{abstract}

\section{Introduction}

Asymptotically flat spacetimes in general relativity are intended to represent ``isolated systems,'' i.e., systems far removed from the influence of other bodies. In order to give a precise definition of  asymptotic flatness, one must specify the precise rate at which the metric approaches a Minkowski metric at asymptotically large distances. There is no algorithm for doing this---i.e., reasonable people may disagree on the precise fall-off conditions that may be used in the definition---but there are two guiding principles that must be respected: (I) The fall-off conditions must not so strong that they exclude the existence of phenomena that could otherwise occur in the deep interior of the spacetime. (II) The fall-off conditions should be sufficiently strong that useful notions that characterize the system, such as total mass and radiated energy flux, are well defined.

Asymptotic flatness conditions have been considered at both spatial infinity (i.e., asymptotically large distances on a Cauchy surface) and null infinity (i.e., asymptotically large distances along null geodesics). The situation at spatial infinity is relatively straightforward in that one need only specify asymptotic conditions on the initial data, since the initial data determines a solution. The positive mass theorem establishes that guiding principle (I) will be violated if one attempts to impose fall off conditions that are so strong as to imply vanishing mass (e.g., fall off faster than $1/r^{d-3}$ in $d$ spacetime dimensions). On the other hand, the Corvino-Schoen gluing theorems~\cite{glue1} (see also \cite{glue2}) establish that guiding principle (II) will hold even if one requires that the initial data agrees exactly with Kerr/Myers-Perry in a neighborhood of infinity\footnote{The gluing theorem 
presented in~\cite{glue1} explicitly only treats the case $d=4$. However, the prerequisite weighted Sobolev inequalities also work in $d>4$~\cite{glue2}, see also appendix~D of~\cite{hollands3}. 
It is then evident from the construction~\cite{glue1} that the gluing theorem continuesthe to hold also in $d>4$ provided 
one has a family of stationary, asymptotically flat metrics whose conserved ADM-quantities exhaust all possible values compatible with the positive mass theorem.
Such a family is provided by the Myers-Perry solutions if we also apply an arbitrary asymptotic boost to these metrics~\cite{Chrusciel}.}. 
Thus, there is ample room for defining asymptotic conditions compatible with both (I) and (II), and the precise choice is largely a matter of taste and convenience. It should be noted that, in general, if the choice of asymptotic fall off conditions is made weaker, then the group of asymptotic symmetries (i.e., the diffeomorphisms that preserve these asymptotic conditions) is made larger. For sufficiently strong fall off conditions at spatial infinity compatible with (I) and (II), the group of asymptotic symmetries will be the Poincare group, whereas for weaker choices one can get enlargements of the Poincare group. In view of the Corvino-Schoen gluing theorems, there is no essential reason not to impose sufficiently strong asymptotic conditions to reduce the group of asymptotic symmetries at spatial infinity to the Poincare group. This conclusion holds in all spacetime dimensions.

The situation at null infinity is considerably less straightforward. The main difficulty here is that one is not really free to specify asymptotic conditions at null infinity; rather one must accept whatever one will get from evolving nonsingular, asymptotically flat initial data on a Cauchy surface (for some suitable notion of asymptotically flat initial data). Thus, in particular, if weak cosmic censorship is false, one would violate guiding principle (I) merely by requiring nonsingular behavior at null infinity. Nevertheless, one can obtain insights into the expected behavior at null infinity from general theorems that hold with small data~\cite{klainermann,rod} as well as from linearized perturbation theory about general solutions~\cite{GX,HI}. These results support the validity\footnote{One can argue about the precise smoothness conditions that should be imposed at null infinity. These are related to the precise choice of asymptotic conditions at spatial infinity and, in our view, are largely a matter of taste and convenience.} of the asymptotic conditions originally proposed by Bondi et al~\cite{bms} and elegantly reformulated in terms of conformal null infinity, $\sI$, by Penrose~\cite{penrose2}. 
In section 2 below, we will review this formulation of asymptotic conditions at null infinity in all even dimensional spacetimes\footnote{In odd dimensions, one has the well-known difficulties in defining $\sI$ \cite{Hollands:2004ac}, so the analysis used in this paper does not apply. See also \cite{Tanabe2009,Tanabe2011} for a different approach.}. 

As is well known and as we shall review in section 2, with the above notion of asymptotic flatness at null infinity, the asymptotic symmetry group at null infinity is an enlargement of the Poincare group known as the BMS group~\cite{bms}. The BMS group contains an infinite dimensional commutative normal subgroup of ``supertranslations.'' It is natural to ask whether this enlargement of the Poincare group is essential or---as is the case with spatial infinity---one could impose stronger asymptotic conditions without violating guiding principle (I) that would reduce the group of asymptotic symmetries at null infinity to the Poincare group. In this paper, we shall argue that the enlargement of the group of asymptotic symmetries to the BMS group is essential in $4$-spacetime dimensions, but in higher even-dimensional spacetimes, stronger asymptotic conditions can naturally be imposed at null infinity that reduce the asymptotic symmetry group to the Poincare group. The reason for this difference is the presence of a ``memory effect'' in $4$ dimensions and---as we shall prove here---its absence in higher even-dimensional spacetimes.

The memory effect is the permanent displacement of an arrangement of freely floating test masses, initially at rest, that is produced by the passage of a burst of gravitational radiation. Here we are concerned only with the displacement occurring at the same $1/r$ order from the source as the usual oscillating displacements caused by gravitational radiation\footnote{There are also lower order in $1/r$ contributions to memory discussed in detail e.g. in \cite{enna}. In particular, there is at lower order the usual acceleration that one even has in static spacetimes such as Schwarzschild.}. It was first discovered in \cite{zeldovich, brag, thorne1} in the context post-Newtonian approximations to full general relativity. Later, Christodoulou~\cite{christodoulu} showed that there is also a contribution to memory arising from gravitational energy fluxes. The relationship between these considerations was clarified in \cite{wise,thorne2}, and more recently~\cite{garfinkle, tolish}. 
For a recent review with many more references, see \cite{garfinkle2}.

In this paper, we shall show that the memory effect is unique to general relativity in four spacetime dimensions, i.e., we shall show that there is no memory effect in higher even dimensions $d > 4$ at the leading $1/r$ order where gravitational radiation causes test masses to accelerate. More precisely, we consider a spacetime satisfying the vacuum Einstein equations near null infinity that is stationary before some advanced time $u_0$ and then again after some later advanced time\footnote{If we restrict consideration to spacetimes that are stationary for $u < u_0$---as can be arranged by a choice of initial data, as discussed above---then one cannot expect the spacetime to become exactly stationary for $u > u_1$, i.e., we should consider a limit as $u_1 \to \infty$. Furthermore, it should be noted if one considers a scattering process wherein one has particle-like matter sources that are incoming from infinity and/or outgoing to infinity at asymptotically early/late times, the spacetime will not be sufficiently stationary near future null infinity at early and late times for the analysis of this paper to apply. However, in a companion paper \cite{GHITW}, we will analyze such scattering processes in the linearized approximation and show that our results on the absence of memory when $d>4$ continue to apply.} $u_1$. We establish that during the radiation epoch $u_0< u < u_1$, the time dependent metric components appear at order $r^{-d/2+1}$ (the precise statement refers to a geometrically constructed coordinate system explained in the body of this paper). On the other hand, the differences between the metric components before and after the radiation epochs are shown to appear at order $r^{-d+3}$. Precisely in $d=4$, these orders are the same. But in higher dimensions, $d/2-1<d-3$, so the leading non-trivial metric components return to their original values after the radiation epoch. It follows that there is no memory effect in higher dimensions.

Our analysis also illuminates further the connection between the memory effect and BMS supertranslations, discussed first in \cite{strominger}. The supertranslations are generated by vector fields of the form $X=T(z) \partial/ \partial u +$ lower order terms, where $T$ is a function of the angular variables $z \in S^{d-2}$. Our analysis shows that the metrics before and after the radiation epochs are related by an asymptotic boost together with such a BMS supertranslation, i.e., we can bring the metrics before and after the radiation epochs into a reference rest frame by a combination of a boost and a BMS supertranslation. The boost going from the frame before to that after the radiation epoch is directly related to the change in Bondi 4-momentum $(\EB, \PB)$. 
In $d=4$ dimensions, we shall show in Prop. 8 below that the supertranslation relating the initial and final metrics is given by 
\ben
\begin{split}
T(\z) =
& 2(\EB - \PB \cdot \z) \log (\EB - \PB \cdot \z) \bigg|_{u_0}^{u_1}\\
& - 16\pi \sum_{l=2}^\infty \frac{1}{ \prod_{s=1}^4 (l-2+s)}  \F^{(l)}_{i_1 \dots i_l} \z^{i_1} \cdots \z^{i_l} \ , 
\end{split}
\een
where $\F(z)$ (see~\eqref{fperangle}) is the flux of gravitational radiation per angle $z \in S^2$ -- identified on the right side with a unit vector $\z \in \mathbb{R}^3$ --
and where $\F^{(l)}_{i_1 ... i_l}$ are its $l$-th multi-pole moments~\eqref{multipole}. 
This formula, which generalizes eq. (3.7) of \cite{strominger}, again relates the supertranslation parameter to physically observable quantities in $d=4$, namely the flux and the Bondi energy-momenta. As previously found by \cite{strominger}, when $d=4$ this supertranslation is related to the ``displacement tensor'' appearing in the memory effect by 
\ben
\Delta_A{}^B = -(D_A D^B T - \frac{1}{2} \delta_A{}^B D^C D_C T) \, ,
\een
where $A,B, \dots$ are tensor indices referring to the angular coordinates. 
However, in $d > 4$, the frames before and after the radiation epochs are only boosted---not  supertranslated---relative to each other. As we shall show in this paper, we have $T(z) = 0$ and $\Delta_A{}^B = 0$ when $d > 4$. 
 
The presence of a memory effect in $d=4$ shows that if we wish to treat frames at early and late time stationary eras on an equal footing, then we must allow supertranslations as asymptotic symmetries, and we cannot impose stronger asymptotic conditions at $\sI$ that would reduce the asymptotic symmetry group to a group smaller than the BMS group. Conversely, the absence of a memory effect in $d > 4$ suggests that for $d>4$ there is no need to allow supertranslations as asymptotic symmetries, and the asymptotic symmetry group at null infinity can be taken to be the Poincare group. This argument may be made more precise as follows. 

In $d$ dimensions, the time dependent metric components encoding radiation start at order $r^{d/2-1}$. In addition, there can be a non-trivial time-independent part that comes in at order $r^{-1}$. In $d > 4$, we may apply a BMS supertranslation to impose, as a gauge condition, the absence of this $r^{-1}$ part. In particular,
if this $r^{-1}$ part is removed in one stationary epoch, then it is also removed in any other. With this gauge condition imposed, the group of allowed asymptotic symmetries is just the Poincare group.
By contrast, in $d=4$, this is not possible, since imposition of such a condition would remove all radiating solutions. 

Of course, one still would be free to not impose such stronger gauge conditions when $d > 4$ and thereby allow supertranslations as asymptotic symmetries in $d>4$, as has been suggested in~\cite{strominger2}. However, there would appear to be little advantage in doing so. In particular, when $d > 4$, the symplectic flux through $\sI$ associated with supertranslations diverges~\cite{HI}. This implies that one cannot define, at least in as far as we can see, a Hamiltonian generator conjugate to a supertranslation in higher dimensions. Thus, even if one were to allow supertranslations, there appears to be no reasonable notion of the ``charge'' or ``flux'' associated with a supertranslation when $d > 4$.

In section \ref{af}, we review the notion of asymptotic flatness at null infinity and the notion of asymptotic BMS symmetries that it gives rise to. In section \ref{s1}, we consider the asymptotic form of the metric near null infinity in the case of a stationary spacetime. The asymptotic form of a non-stationary metric near null infinity is then obtained in section \ref{s2}. The relationship between the form of the metric before and after a burst of gravitational radiation is analyzed in section \ref{memory}. In $d=4$, this difference is characterized by an asymptotic boost and a supertranslation, but for $d > 4$, the supertranslation vanishes. We show that the memory effect is directly related to the supertranslation characterizing the difference between the metric before and after the radiation burst. Thus, the memory effect vanishes when $d > 4$. Our reasons for concluding that the asymptotic symmetry group at null infinity should be taken to be the Poincare group when $d > 4$ are summarized in section \ref{summary}.

In a companion paper \cite{GHITW}, we will analyze the memory effect in linearized gravity off of Minkowski spacetime for point particles undergoing a local interaction. We will show there by explicit calculation that the memory effect vanishes when $d > 4$. 

\medskip
\noindent
Our {\bf conventions and notations} for the signature, Riemann curvature tensor etc. are the same as in~\cite{wald}. $d$ is the dimension of the spacetime $\M$. 
Bold face letters such as $\PB$ or $\z$ refer to vectors in $\mathbb{R}^{d-1}$, and the standard inner product between such vectors is denoted by a dot. Our units are such that $G_N=1$. 

\section{Asymptotic Flatness and Asymptotic Symmetries at Null Infinity}\label{af}

An even dimensional ($d \ge 4$) spacetime $(\M,g)$ is said to be asymptotically flat at null infinity  
if the following conditions are satisfied:
\begin{enumerate}
\item[(i)] There exists a function $\Omega$ 
such that $\tilde g = \Omega^2 g$ can be smoothly extended to an ``unphysical'' spacetime $\tilde \M$ with boundary $\partial \tilde \M \supset \sI^+ \cup \sI^-$ and 
$\sI^\pm \cong \Sigma \times {\mathbb R}$, 
where $\Sigma$ is a compact $(d-2)$ dimensional manifold. 

\item[(ii)]
$\tilde n = {\rm grad}_{\tilde g} \Omega$ is a null vector field on $\sI^+$ relative to $\tilde g$, and on $\sI^\pm$, there holds $\d \Omega \neq 0$. 

\item[(iii)] The vacuum Einstein equations $Ric_g=0$ hold for $g$ in an open neighborhood of $\sI^+$. 
\end{enumerate}

For a more detailed discussion, see~\cite{G} or chapt.~11 of~\cite{wald}.
A characterization of precisely which initial data sets yield a time evolved metric that will satisfy these asymptotic conditions is an extremely difficult 
dynamical problem. As already mentioned in the Introduction, a good indication supporting the asymptotic flatness conditions is that they are
preserved by smooth linear perturbations that have compact support in the interior of a Cauchy surface, see~\cite{GX} for $d=4$ and~\cite{HI} in $d>4$. 
Semi-global results (evolving data backwards from $\sI^+$) in the non-linear regime were obtained in \cite{Anderson} (in all even $d>4$) following the pioneering work of \cite{Friederich}
(in $d=4$)~\footnote{These results indicate that in $d>4$, more stringent conditions can be imposed than expressed by (i)-(iii). We will see that these conditions are 
actually a consequence, see Thm.~\eqref{thm2}.}. 
Global results which establish (i)--(iii) for small but non-linear perturbations of Minkowski are~\cite{klainermann} (in $d=4$)
and \cite{rod} (for all even $d>4$). 

In this paper, we will simply assume (i)--(iii). It is then always possible to choose the conformal factor $\Omega$ in such a way such that the metric $g$ has ``conformal Gaussian null form'' near $\sI^+$. 
By definition, this means that
\ben\label{Pen}
g = \frac{1}{\Omega^2} \Bigg\{ 2 \d u(\d\Omega - \Omega^2 \alpha \d u- \Omega \beta_A \d z^A) + \ga_{AB} \d z^A \d z^B \Bigg\} = \frac{1}{\Omega^2} \tilde g \ , 
\een
where  $\alpha, \beta, \ga$  can be viewed as tensor fields $\Sigma$ that smoothly depend on $\Omega,u$, and where $z^A$ are local coordinates on $\Sigma$. 

The name arises first of all because the unphysical metric $\tilde g$ is in Gaussian null form\footnote{Gaussian null coordinates were introduced in~\cite{Penrose}.}.
Geometrically these coordinates have the following properties. On $\sI^+$, $\tilde n = \partial/\partial u$ is tangent to affinely parameterized null geodesics ruling $\sI^+$ relative to $\tilde g$. The coordinates $z^A$ are defined first on a cross-section, $\Sigma(u_0)$,
corresponding to the constant value 
$u_0$ of the coordinate $u$, and then they are transported to all of $\sI^+$ along $\tilde n$.
The vector field $\tilde l=\partial/\partial \Omega$ is another affinely parameterized (relative to $\tilde g$) null vector field transverse to $\sI^+$ which is normalized 
so that $\tilde g(\tilde n, \tilde l)=1$, and such that it is orthogonal to the
cross sections $\Sigma(u)$ on $\sI^+$. The coordinates $(u, z^A)$ are then transported to 
a sufficiently small neighborhood of $\sI^+$ by transport along $\tilde l$. The term ``conformal'' refers to the key point  that 
$\Omega$ is not only an affine parameter of null geodesics of $\tilde g$, but also coincides with the conformal factor. 

If we define
\ben
r=\frac{1}{\Omega}, 
\een
we get from~\eqref{Pen} a ``Bondi-type''\footnote{Note that this coordinate system 
actually differs from what is normally called ``Bondi gauge'', although it has obvious similarities. See, e.g., \cite{Ishibashi:2007kb}.} coordinate system:
\ben\label{Pen1}
g =  -2 \d u(\d r + \alpha \d u+ r \beta_A \d z^A) + r^2 \ga_{AB} \d z^A \d z^B  \ , 
\een
and this is just another way to state our asymptotic conditions. The coordinate vector field $\partial/\partial r$ defined by 
the system $(r,u,z^A)$ is easily checked to be tangent to affinely parameterized null geodesics relative to $g$. This fact is ultimately responsible 
for the preferred geometrical status of our coordinate system.

For $d$-dimensional Minkowski spacetime $(\M=\mathbb{R}^d, g=\eta)$, we have $\Sigma \cong S^{d-2}$, $(u,r,z^A)$ are given by $u=t-r, r=|\x|, z^A = {\rm angles}$, and in fact
\ben\label{eta}
\eta = -2 \d u(\d r + \tfrac{1}{2} \d u) + r^2 s_{AB} \d z^A \d z^B
\een
where here and in the rest of the paper $s_{AB}$ denotes 
the unit radius round sphere metric. For Schwarzschild, $u=t-r_*$ with $r_*$ given by the tortoise coordinate. 

It can be readily seen that the remaining freedom in specifying a conformal 
Gaussian null coordinate system is as follows: We can make a different choice of the initial 
cross section $\Sigma(u_0)$; we can change the affine coordinate $u$ on $\sI^+$ by an affine transformation for each null generator of $\sI^+$; we can similarly 
apply a suitable ``angle-dependent'' affine transformation of the 
parameter $r$ consistent with  the requirement $g(\partial/\partial u, \partial/\partial r)=1$ on $\sI^+$; we can apply a diffeomorphisms to the points $z$ 
in $\Sigma(u_0)$. Thus our freedom consist of:
\begin{enumerate}
\item Changing $r \mapsto r+S(u,z)$, 
\item Changing $u \mapsto e^{\omega(z)} u$ on $\sI^+$ and making corresponding changes to $r,z^A$,
\item Changing $u \mapsto u + T(z)$ on $\sI^+$ and making corresponding changes to $r,z^A$,
\item Applying a diffeomorphism to points $z \in \Sigma(u_0)$. 
\end{enumerate}
It is easy to see that by choosing $\partial_u S$ in the transformation of type 1) suitably, we can 
set $\alpha$ to a constant on $\sI^+$, which we assume has been done. For convenience, we assume that 
the constant is $\alpha = 1/2$ on $\sI^+$, as in Minkowski spacetime~\eqref{eta}. Thus, the remaining transformations of type 1) are 
ones where $S$ does not depend on $u$. 

A transformation of type 2) changes the induced metric $\ga_{AB}$ on $\Sigma$ by the conformal factor $e^{2\omega}$. 
In $d=4$, one typically assumes that, topologically $\Sigma \cong S^2$, for otherwise one could clearly not say that the metric is asymptotically 
Minkowskian. We can then turn $\ga_{AB}$ into the metric $s_{AB}$ on the unit round sphere on $\sI^+$ for one particular cross-section of $\sI^+$, say $\Sigma(u=0)$. In higher dimensions, we  assume as a strengthening of our asymptotic conditions:
\begin{enumerate}
\item[(iv)]
 $(\Sigma, \ga_{AB})$ is conformal to an Einstein space for some value of $u$ (we do not have to do this, but then it seems impossible to define Bondi mass, news etc., see~\cite{thorne}). In fact, we will assume that this Einstein space is a unit round sphere $(S^{d-2}, s_{AB})$\footnote{Einstein's equation relate the value of $\alpha$ 
on $\sI^+$ to the scalar curvature of $\ga_{AB}$ on $\sI^+$. We have anticipated this relation in our choice for the normalization.}. 
\end{enumerate} 
Einstein's equations 
then imply~\cite{thorne} that $\ga_{AB}$ is equal to $s_{AB}$ on $\sI^+$ for {\em any} value of $u$.
Thus, we have fixed the first two ambiguities. 

The ambiguity of type 3) corresponds to the 
choice of the cross section distinguished by $u=u_0$, i.e. a change in the cross section. This corresponds to BMS supertranslations. 

The last ambiguity type 4) is cut down because we have already
demanded that, on $\sI^+$, $\ga_{AB}$ is the metric of the round sphere, $s_{AB}$. So we are left with diffeomorphisms that are conformal transformations of this round sphere (group 
${\rm O}(d-1,1)$), and then we must combine such a diffeomorphism with a transformation of type 2) to compensate the conformal factor.  

Summarizing our discussion, we can say that the remaining diffeomorphisms of the metric respecting our gauge choices near $\sI^+$ are generated by linear combinations of the following vector fields $X$ defined in a neighborhood of $\sI^+$:

\begin{enumerate}
\item[(I)] Vector fields 
\ben
X=S(z) \frac{\partial}{\partial r} + \dots \ , 
\een
where $S$ is an arbitrary smooth function on $S^{d-2}$.

\item[(IIa)] Vector fields 
\ben\label{super}
X = T(z) \left( \frac{\partial}{\partial u} - \frac{\partial}{\partial r} \right) - \frac{1}{r} s^{AB} D_B T(z) \frac{\partial}{\partial z^A} + \dots \ ,
\een 
where $T$ is any smooth function on $S^{d-2}$ orthogonal to the $l=0,1$ spherical harmonics. 
These $X$ are precisely the ``BMS supertranslations''. 

\item[(IIb)]
Vector fields $X$  of the form~\eqref{super}, where $T$ is a linear combination of $l=1$ spherical harmonics on $S^{d-2}$. 
Such $T$ satisfy 
\ben
D_A D_B T = -s_{AB} T.
\een
These $X$
correspond to infinitesimal spatial translations of the underlying ``Minkowski spacetime'' given by eq.~\eqref{eta}. 

\item[(IIc)] The vector field
\ben
X = \frac{\partial}{\partial u} + \dots \ . 
\een
This $X$ corresponds to an infinitesimal time translation. 

\item[(III)] 
Vector fields 
\ben
X = C(z)\left[ u \left( \frac{\partial}{\partial u} - \frac{\partial}{\partial r} \right) - r\frac{\partial}{\partial r} \right] - s^{AB} D_B C(z) \frac{\partial}{\partial z^A} + \dots \ , 
\een
where $C$ is an $l=1$ spherical harmonic--implying that 
$s^{AB} D_B C = \xi^A$ is a conformal Killing vector field of $S^{d-2}$. These $X$ correspond to infinitesimal Lorentz boosts. 

\item[(IV)] 
Vector fields of the form 
\ben
X = \xi^A \frac{\partial}{\partial z^A} + \dots \ , 
\een 
where $\xi^A$ is an isometry of $S^{d-2}$. These $X$ correspond to infinitesimal rotations. 
\end{enumerate}

In all cases the dots $\dots$ stand for a vector field that vanishes on $\sI^+$ together 
with its first and second derivative (relative to an arbitrary derivative operator on $\tilde \M$ such as $\tilde \nabla$). 
This vector field is uniquely determined in each case by the requirement that $\pounds_X g$ satisfies the ``linearized form'' of our gauge conditions~\eqref{Pen1}, i.e. 
has vanishing $(ru), (rA), (rr)$ components. Thus the parts of $X$ indicated by dots depend in general on $g$ (but happen to vanish for $\eta$), but the leading terms displayed in (IIa,b,c)--(IV) manifestly do not. 

The vector field of type (I) is to be viewed as a ``gauge transformation'', since it vanishes on $\sI^+$. The others (IIa,b,c)--(IV) generate the Lie algebra of the 
``asymptotic symmetry group'', which is called the BMS group. More precisely, the BMS Lie algebra is isomorphic to
\ben
\frak{bms}_d = \frac{{\rm span} \{X \in \text{(IIa,b,c),(III),(IV)} \}}{ \{ X \text{vanishing to second order at $\sI^+$} \}} \ . 
\een
The vector fields (IIa,b,c) correspond to an infinite-dimensional abelian normal Lie sub algebra, $\frak{t}_d$, and the quotient 
$\frak{bms}_d/\frak{t}_d \cong \frak{so}(d-1,1)$ is the Lorentz-Lie algebra. It can be identified (non-canonically) with the Lie sub algebra corresponding to the 
vector fields (III),(IV). An asymptotic symmetry $\phi$ generated by one of these vector fields acts on the cross sections $\cong S^{d-2}$ of $\sI^+$ as a conformal 
transformation, called $\hat \phi$, so $\hat \phi^* s_{AB} = e^{2\omega} s_{AB}$. 
For an asymptotic boost with velocity parameter $\v$ [see~footnote 10 below], 
the action and conformal factor are concretely:
\ben\label{conf}
\hat \phi: \z \mapsto \frac{\z \sqrt{1-v^2} + \v [(\v \cdot \z)/(1+\sqrt{1-v^2}) -1]}{1-\v \cdot \z} \ , \quad \omega(\z) =  \log \frac{\sqrt{1-v^2}}{1-\v \cdot \z} , 
\een
where here and in the following, points $z$ in $S^{d-2}$ are identified with unit vectors $\z \in \mathbb{R}^{d-1}$. If $X \in \frak{t}_d$ is a vector field corresponding to an 
infinitesimal symmetry of type (IIa,b,c) with function $T(z)$, then the element $\hat X = \phi^* X \in \frak{t}_d$ conjugate under 
$\phi$ corresponds to $\hat T(z) = e^{-\omega(z)} T(\hat \phi(z))$. 

Later we will see that for $d>4$, it is possible to impose more stringent conditions on the metric, which will imply that the true asymptotic symmetry algebra is 
reduced to the Poincar\'e Lie algebra $\frak{p}_d = \frak{so}(d-1,1) \ltimes {\mathbb R}^d$ which is the Lie sub-algebra of $\frak{bms}_d$ excluding 
supertranslations (IIa). 

\medskip

In the following two sections, we analyze systematically the consequences of the vacuum Einstein equation $Ric_g=0$ near $\sI^+$. It is convenient to make 
the Taylor expansions 
\ben\label{coeff}
\ga_{AB} \sim \sum_{n \ge 0} \ga^{(n)}_{AB} \ r^{-n} \ , \quad 
\beta_{A} \sim \sum_{n \ge 0} \beta^{(n)}_{A} \ r^{-n} \ , \quad 
\alpha \sim \sum_{n \ge 0} \alpha^{(n)} \ r^{-n}
\een
where $\sim$ indicates that these expansions might not be convergent. 
Each of the ``coefficients'' $\ga_{AB}^{(n)}, \beta_A^{(n)}, \alpha^{(n)}_{}$ is a tensor field on 
$S^{d-2}$ depending on $u$. 
Einstein's equations then give relationships between the coefficients. 
For instance, we have, in any even dimension $d\ge 4$ 
\ben\label{initial}
\ga^{(0)}_{AB} = s^{}_{AB} \ , \quad \alpha^{(0)}_{} = 1/2 \ , \quad \beta^{(0)}_A = 0 \ . 
\een
The first two conditions are in fact, as already discussed, consequences of our asymptotic conditions/gauge choices, but the last one is one such consequence of the Einstein equations~\cite{thorne}. These zeroth-order relations serve as ``initial conditions'' to constrain the subsequent orders via Einstein's equations. In fact, 
these consequences where worked out in general dimension already in~\cite{thorne}. 
These will be recalled in sec.~\ref{s2}. 
We first restrict attention to {\em stationary} metrics, 
where more stringent conclusions can be drawn. 

\section{Stationary Metrics}\label{s1}

We ask what we additionally learn from Einstein's equations $Ric_g =0$ near $\sI^+$ if we demand that $g$  is stationary in a neighborhood of $\sI^+$.
By this we mean that there exists a Killing field $K$, 
\ben\label{Killing}
\pounds_K g = 0 \ , 
\een
with time-like orbits near $\sI^+$.  
Any Killing field is a forteriori an asymptotic symmetry of the metric, so it must be given by a linear combination of the vector fields $X$ in items (I)-(IV). 
Since $K$ is time-like near infinity, the boost- (III) and rotation- (IV) parts must be absent, so 
\ben
K = \frac{\partial}{\partial u} + T(z) \left( \frac{\partial}{\partial u} - \frac{\partial}{\partial r} \right) - \frac{1}{r} s^{AB} D_B T(z) \frac{\partial}{\partial z^A} +
S(z) \frac{\partial}{\partial r} + \dots
\een
up to an overall positive constant. Here dots represent a vector field whose derivatives up to second order vanish at $\sI^+$, and $T$ must satisfy $|T(z)|<1$, again since 
$K$ is timelike near $\sI^+$. Without loss of generality, we can assume that $T$ is orthogonal to the constant function on $S^{d-2}$. 
We now wish to argue that $K$ is actually a linear combination of a time-translation and 
a spatial translation, i.e. that $T$ is a linear combination of the $(d-1)$ $l=1$ spherical harmonics on $S^{d-2}$, and $S=0$. 
We start by writing out Killing's equation~\eqref{Killing} in CGNCs. We find from the $(AB),(Au),(uu)$-components respectively
\ben
\begin{split}\label{kill}
0&=(1+T) \partial_u \ga^{(1)}_{AB} - 2(D_A^{} D_B^{} + s_{AB}^{})T + 2s_{AB} S \ ,  \\
0&=(1+T) \partial_u \beta_A^{(1)} + D_A^{} S \ , \\
0&=(1+T) \partial_u \alpha^{(1)}_{} \ . 
\end{split}
\een 
The other components of Killing's equation do not give further information but determine the higher order terms represented by 
dots in our formula of $K$. 
To make progress, we must also use Einstein's equations. The analysis is rather different 
in $d=4$ respectively higher even dimensions, so we treat both cases separately. 

\medskip
\noindent
{\bf Dimension $d>4$ (and even):} \\
Using our ``initial conditions''~\eqref{initial}, the $(AB)$ components of Einstein's equations give
\ben\label{AB0}
0=(d-4) \partial_u \ga^{(1)}_{AB} + s_{AB} \partial_u \ga^{(1)}_{} \ , 
\een
where here and in the following, $\ga^{(1)}=s^{AB} \ga^{(1)}_{AB}$. This immediately gives $\partial_u \ga^{(1)}_{AB}=0$. 
The $(rA)$ components of Einstein's equations give
\ben\label{rA0}
0 = D_{[A} \ga^{(1)A}{}_{B]}  +  (d-3) \beta^{(1)}_B \ ,  
\een
implying $\partial_u \beta^{(1)}_A = 0$, too. The second equation in~\eqref{kill} then gives $D_A S = 0$, so $S$ must be 
constant. The trace of the first equation in~\eqref{kill} furthermore gives $(d-2) S = (D^C D_C+ d-2)T$. Since $T$ is orthogonal to 
the constant function on $S^{d-2}$, it follows that $S=0$, and then it follows that $T$ is a linear combination of $l=1$ spherical harmonics, 
as desired. 

\medskip
\noindent
{\bf Dimension $d=4$:} \\
In this case, \eqref{AB0} only allows us to conclude that $\partial_u \ga^{(1)}=0$, which using the first equation in~\eqref{kill} gives us 
$S = \frac{1}{2}(D^C D_C+ 2)T$. From the third equation in~\eqref{kill} we get $\partial_u \alpha^{(1)}=0$, since $|T(z)|<1$. The first 
two equations in~\eqref{kill} become
\ben
\begin{split}\label{kill1}
0&= (1+T) \partial_u \ga^{(1)}_{AB} - 2(D_A^{} D_B^{} - \tfrac{1}{2} s_{AB}^{} D^C D_C)T  \ ,  \\
0&=(1+T) \partial_u \beta_A^{(1)} + \tfrac{1}{2} D_A^{} (D^C D_C+ 2)T \ . 
\end{split}
\een 
Taking 
a $u$-derivative of~\eqref{rA0} furthermore gives 
\ben\label{uA0}
\partial_u \beta^{(1)}_A = - \frac{1}{2} D^B \partial_u \ga^{(1)}_{AB} \ ,  
\een
and applying $D^B$ to the first equation in~\eqref{kill1}, and then substituting the second equation in~\eqref{kill1} in order to 
eliminate $\partial_u \beta^{(1)}_A$ then also gives $(\partial_u \ga^{(1)}_{AB}) D^B T = 0$. Since $T$ does not depend on $u$, 
the first equation in~\eqref{kill1} also gives $\partial_u^2 \ga^{(1)}_{AB}=0$. We next combine this information with 
the $(uu)$ and $(ru)$ components of Einstein's equation (see~\eqref{below-1}), 
which yields
\ben\label{ru0}
D^A \partial_u \beta^{(1)}_A = 0 \ . 
\een
Taking $D^A$ of the second equation in~\eqref{kill1} and using~\eqref{ru0} gives the first equation in
\ben
\begin{split}\label{kill2}
0&=(D^AT) \partial_u \beta^{(1)}_{A} + \tfrac{1}{2}D^B_{} D_B^{} (D_A^{} D^A_{} +2)T  \\
0&=(D^AT) \partial_u \beta^{(1)}_{A} - \tfrac{1}{2}(D^A_{} D^B_{} T) \partial_u \ga^{(1)}_{AB} \ ,  
\end{split}
\een
while the second equation is obtained by taking $D^A$ of $(\partial_u \ga^{(1)}_{AB}) D^B T = 0$ and 
using~\eqref{uA0}. Subtracting the second equation from the first, using the first equation in \eqref{kill1} to 
eliminate $\partial_u \ga^{(1)}_{AB}$, and integrating the result over $S^2$ gives 
\ben
\int_{S^2} \frac{(D^A D^B T)D_A D_B T - \frac{1}{2} (D^A D_A T)^2 }{1+T} = - \frac{1}{2}  \int_{S^2} D^B_{} D_B^{} (D_A^{} D^A_{} +2)T = 0 \ . 
\een 
Now let $A^A{}_B = D^A D_B T$. This is a self-adjoint map in the tangent space of each point of $S^2$ (with respect to the 
inner product $s_{AB}$), and so has two real eigenvalues $\lambda_1, \lambda_2$. In terms of the eigenvalues, the 
integrand on the left side becomes $\frac{1}{2}(\lambda_1-\lambda_2)^2/(1+T)$, whereas the right side vanishes by Gauss' theorem. 
Since $|T(z)|<1$, we therefore conclude 
that $\lambda_1=\lambda_2$, so $D^A D_B T = \lambda_1 \delta^A{}_B$ for some function $\lambda_1$. Lowering the index $A$ 
it follows that $D_A D_B T = \frac{1}{2} s_{AB} D^C D_C T$, meaning that $T$ must be an $l=1$ spherical harmonic, and then $S$ has to vanish. 
This concludes the argument in $d=4$.  

\medskip
\noindent
If we want, we can identify points on $S^{d-2}$ with unit vectors ${\bf z}$ in ${\mathbb R}^{d-1}$, and then we can write
$T(z) = {\bf v} \cdot {\bf z}$. Since $|T(z)|<1$, we must have $|{\bf v}|<1$. Thus, $K$ must be an ordinary asymptotic infinitesimal translation into a time-like direction. 
Let $\phi$ be the unique (near $\sI^+$) asymptotic boost with velocity parameter $-{\bf v}$~\footnote{The boost is explicitly 
\ben \nonumber 
\Lambda = \left( 
\begin{matrix}
\frac{1}{\sqrt{1-v^2}} & \frac{-\v}{\sqrt{1-v^2}}\\
\frac{-\v}{\sqrt{1-v^2}} & I + v^{-2}\left( \frac{1}{\sqrt{1-v^2}}-1 \right) \v \otimes \v
\end{matrix}
\right) ,
\een
and it acts on $(t, \x)$, where $t=r+u$ and $\x = r\z$. 
}. Then 
$
\phi^* K \propto \partial/\partial u \ , 
$
at $\sI^+$, which is a Killing field for $\phi^* g$. For convenience of notation, we now denote $\phi^* K, \phi^*g$ again by $K,g$. Then we can say that
\ben\label{Kdef}
K = \frac{\partial}{\partial u}
\een
at $\sI^+$. 
We claim that this formula must hold not only at $\sI^+$ but in a full neighborhood. 
This can be seen as follows. Let $\psi_t$ be the 1-parameter group of diffeomorphisms generated by $K$. By construction, it acts on points 
of $\sI^+$ by $u \mapsto u+t, z \mapsto z$. Since each $\psi_t$ is an isometry of $g$, it follows that it preserves affinely parameterized geodesics. So it must act on $r$ by an affine transformation and leave $z^A$ invariant off $\sI^+$, i.e. we can say that points labelled by $(u,r,z^A)$ 
get mapped to points labelled by $(u+t, e^{\omega_t(z,u)}r + S_t(z,u), z^A)$ for sufficiently small $r$. This action is incompatible with the form of our asymptotic 
symmetries (I)-(IV) unless $S_t=\omega_t=0$, and it then follows immediately that the formula~\eqref{Kdef} holds in a neighborhood of 
$\sI^+$. Hence, since $K$ is Killing for $g$, we conclude that the expansion coefficients of $g$ satisfy 
\ben\label{stat}
\partial_u \ga_{AB}^{(n)} = \partial_u \beta_A^{(n)}=\partial_u \alpha^{(n)}_{} = 0
\een
for all $n \ge 0$. We summarize our findings so far in a lemma:

\begin{lemma}
Let $(\M,g)$ be an asymptotically flat spacetime presented in CGN gauge~\eqref{Pen1}, 
with Killing field $K$ that is timelike near $\sI^+$. Then there exists a unique diffeomorphism $\phi$ defined near $\sI^+$ which is 
an asymptotic boost  and preserves the CGN gauge, such that the expansion coefficients~\eqref{coeff} of $\phi^* g$ are all independent of $u$, and 
such that $\phi^* K=\partial/\partial u$ near $\sI^+$. 
\end{lemma}

We next use the information provided by the lemma in Einstein's equation. Again we denote $\phi^* g$ by $g$ to simplify our notation. 
First, we consider Einstein's equations at order $n=0$ in~\eqref{coeff},  using~\eqref{stat}.
We find from the $(AB)$-components 
\ben\label{ein1}
\begin{split}
0=& -4(d-4) \alpha^{(1)}s_{AB} + 2(d-4) D^{}_{(A} \beta_{B)}^{(1)} + 2s_{AB}^{} D^C \beta^{(1)}_C \\
& + (d-2) \ga^{(1)}_{AB} - s_{AB}^{} \ga^{(1)}_{} - D^C D_C  \ga^{(1)}_{AB} + 2D^{}_{(A} D^C \ga^{(1)}_{B)C} - D^{}_A D^{}_B \ga^{(1)}_{} \ , 
\end{split}
\een
from $(uA)$-components
\ben\label{ein2}
0= 2(d-5) D^{}_A \alpha^{(1)}_{} - D^C(D^{}_A \beta^{(1)}_C - D^{}_C \beta^{(1)}_A ) \ , 
\een
and from the $(ru)$-component
\ben\label{ein3}
0= -2(d-4) \alpha^{(1)}_{} \ .
\een
In order to evaluate the consequences of these equations, the following lemma~\cite{IW} will be rather useful:

\begin{lemma}\label{decomp}
Any smooth 1-form field $v_A$ on $S^{d-2}$ can be decomposed uniquely as 
\ben
v_A = V_A + D_A S \ , 
\een
where $V_A$ is a divergence free 1-form field, $D^A V_A=0$. We refer to $S$ as the ``scalar part'' and $V_A$ as the ``vector part''. 
Any smooth symmetric rank-2 tensor field $t_{AB}$ on $S^{d-2}$ can be decomposed uniquely as
\ben
t_{AB} = T_{AB} + D_{(A} W_{B)} + (D_A D_B -\tfrac{1}{d-2} s_{AB} D^C D_C)T + \tfrac{1}{d-2} s_{AB} U \ , 
\een
where $T_{AB}$ is trace-free and divergence free, $0=D^A T_{AB} = s_{AB} T^{AB}$, and where $W_A$ is divergence free. 
We refer to $T_{AB}$ as the ``tensor part'', $W_A$ as the ``vector part'' and $(T,U=s^{AB} t_{AB})$ as the ``scalar part''. 
\end{lemma}

Some special things happen in low dimension. On $S^2$, the tensor part necessarily vanishes, as there are no divergence free, trace free symmetric rank two tensors. 
Furthermore, the vector part may always be written as $V_A = \epsilon_{AB} D^B V$ for some function $V$. Furthermore, when $d>4$, the scalar, vector, and tensor 
parts on $S^{d-2}$ form inequivalent, irreducible representation spaces of the group ${\rm SO}(d-1)$. The spectrum of the Laplacian $-D^A D_A$ consists of the eigenvalues shown in table~1. 

\begin{table}[h!]
\label{tab:spectrum}
  \begin{center}
      \begin{tabular}{c|c}
Tensor type & Spectrum \\
\hline \hline
scalar & $\{ l(l+d-3) \ : \quad l=0,1,2,\dots \}$ \\
vector & $\{ l(l+d-3)-1 \ : \quad  l=1,2,3, \dots \}$\\
tensor & $\{ l(l+d-3)-2 \ : \quad  l=2,3,4, \dots \}$\\
  \end{tabular}
  \caption{Spectrum of Laplacian $-D^A D_A$ on $S^{d-2}$.}
  \end{center}
\end{table}

As a consequence of the uniqueness of the decomposition, and the inequivalence of the representations, in $d>4$, 
any ${\rm SO}(d-1)$ invariant partial differential operator, such as the Laplacian, leaves the scalar, vector, and tensor subspaces invariant. By contrast, 
on $S^2$ (i.e. for $d=4$), there is no tensor part, and the vector and scalar parts form equivalent representations of ${\rm SO}(3)$. As representations of ${\rm O}(3)$, they 
are inequivalent and correspond to ``parity even'' (polar) and ``parity odd'' (axial) scalar fields on $S^2$.

Armed with this information, we can now easily understand the consequences of eqs.~\eqref{ein1}-\eqref{ein3}. Because there is a difference between the case $d=4$ and the cases
$d>4$ and even, we look at each of them separately. 

\medskip
\noindent
{\bf Dimension $d>4$ (and even):} \\
Eq.~\eqref{ein3} clearly gives $\alpha^{(1)}=0$. Next, we split 
$\beta^{(1)}_A = V_A + D_A S$ into a scalar and a vector part. Eq.~\eqref{ein2} does not give any constraint on the scalar part, and 
gives $D^C D_C V_A = (d-3) V_A$ for the vector part. Given the spectrum of the Laplacian in the vector case, this equation only has the trivial solution $V_A=0$. 
We next decompose eq.~\eqref{ein1} and $\ga^{(1)}_{AB}$ into its scalar, vector, and tensor part. These equations decouple since the differential operators are clearly 
${\rm O}(d-1)$-invariant. For the tensor part, $T_{AB}$, of $\ga^{(1)}_{AB}$ we infer that $D^C D_C T_{AB} = (d-2) T_{AB}$. Given the spectrum of the Laplacian in the tensor 
case, this equation only has the trivial solution. Thus, the tensor part of $\ga^{(1)}_{AB}$ vanishes. Next, for the vector part $W_A$, of $\ga^{(1)}_{AB}$ we get 
$2(d-4) D_{(A} W_{B)} = 0$, so there is no vector part either. Taking a trace of~\eqref{ein1}, we learn for the scalar part $(T,U)$ of  $\ga^{(1)}_{AB}$ that
\ben
D^A D_A \bigg\{ U- (D^B D_B + d-2)T - 2(d-2)S \bigg\} = 0 \ , 
\een
where $S$ is 
the scalar part of $\beta^{(1)}_A$. It follows that the expression in curly bracket is constant, which we can absorb in a constant shift of $S$. 
Thus, all in all we learn that eqs.~\eqref{ein1}-\eqref{ein3} are satisfied if and only if 
\ben\label{eincon1}
\ga^{(1)}_{AB} = (D_A D_B + s_{AB})T + 2Ss_{AB} \ , \quad \beta^{(1)}_A = D_A^{} S \ , \quad \alpha^{(1)}_{} = 0 \ , 
\een 
where $S,T$ are arbitrary smooth functions on $S^{d-2}$. Furthermore, it is checked that, at this order, all other components of Einstein's equation are also satisfied by these expressions. 

\medskip
\noindent
{\bf Dimension $d=4$:} \\
Eq.~\eqref{ein3} gives no constraints on $\alpha^{(1)}$. We next consider eq.~\eqref{ein2}. On $S^2$, the decomposition into scalar 
and vector parts is $\beta^{(1)}_A = D_A S + \epsilon_{AB} D^B V$. We  consider this decomposition  in~\eqref{ein2}. $S$ drops out of the equation 
right away, and taking a divergence, $V$ drops out, too. We are left with $D^C D_C \alpha^{(1)}=0$, which implies that $\alpha^{(1)} \equiv c$ is a constant. Once this is known, 
we learn that $D_A D^C D_C V=0$, hence that $D^C D_C V$ is a constant. Integrating this term over $S^2$ and applying Gauss' theorem shows that the constant in fact vanishes, so 
$V$ must be constant and we learn that $\beta^{(1)}_A$ has no vector part. Consideration of the vector part of 
\eqref{rA0} shows that the vector part $W_A$ of $\ga^{(1)}_{AB}$ obeys $D^A D_{(A} W_{B)}=0$. Multiplying with $W^B$, integrating over $S^2$ and 
integrating by parts gives $D_{(A} W_{B)}=0$, so $\ga^{(1)}_{AB}$ has no vector part.
The rest of the argument is unchanged compared to the case $d>4$, because $\alpha^{(1)}$ drops out of 
\eqref{ein1} (except that $\ga^{(1)}_{AB}$ cannot have a tensor part on $S^2$ in the first place). 

Thus, all in all we learn that eqs.~\eqref{ein1}-\eqref{ein3} are satisfied iff
\ben\label{eincon2}
\ga^{(1)}_{AB} = (D_A D_B + s_{AB})T + 2Ss_{AB} \ , \quad \beta^{(1)}_A = D_A^{} S \ , \quad \alpha^{(1)}_{} = c \ , 
\een 
where $S,T$ are arbitrary smooth functions on $S^2$, and where $c$ is a constant. 
Furthermore, it is checked again that all other components of Einstein's equation are satisfied by these expressions. 

\medskip

We note that $\ga^{(1)}_{AB}, \beta^{(1)}_A$ in eq.~\eqref{eincon1} are ``pure gauge'', and in fact correspond precisely to transformations of the type (I) and (IIa) in the above list
with the same function $S$ and $T$ as in~\eqref{eincon1} resp.~\eqref{eincon2} up to the trivial change $T \to -2T$. 
Furthermore, since the vector fields $X$ of type (I) and (II) commute with $K$, we arrive at the following conclusion:

\begin{lemma}
There exists a diffeomorphism $\psi$ of $\M$ that is an asymptotic symmetry and preserves the CGN gauge~\eqref{Pen1} generated by a linear combination of the 
vector fields $X$ in case (I) and (IIa) such that the expansion coefficients~\eqref{coeff} of $\psi^* g$ have $\ga^{(1)}_{AB}=\beta^{(1)}_A=0$. In $d > 4$ we have additionally $\alpha^{(1)}_{}=0$, and in $d=4$, $\alpha^{(1)} = c$ is constant. Furthermore, $\psi^* K = \partial/\partial u$. 
\end{lemma}

This finishes our analysis in $d=4$, but in $d>4$, we can go further and derive constraints on higher expansion orders~\eqref{coeff} which we will do now. 
To simplify the discussion, we can thus pass from $g$ to $\psi^* g$, which we will do for the rest of this section. We denote this new metric again by $g$ to simplify the notation. 
We claim that we have 
\ben
\ga^{(n)}_{AB} = 0 \ , \quad \beta^{(n)}_A = 0 \ , \quad \alpha^{(n)}_{} = 0\ , 
\een 
where $1\le n \le d-3$ in the first two equations and $1\le n < d-3$ in the last expression, while $\alpha^{(d-3)}=c$ is a constant. To prove this, we proceed by induction. The case 
$n=1$ 
is already settled on account of the previous lemma. Let the statement thus be assumed to be true up to and including order $n$. Then, at order $n+1$, we get from the $(AB)$ components of Einstein's equations 
\ben\label{einn1}
\begin{split}
0=& -4(d-4-n) \alpha^{(n+1)}s_{AB} + 2(d-4-n) D^{}_{(A} \beta_{B)}^{(n+1)} + 2s_{AB}^{} D^C \beta^{(n+1)}_C \\
& + [n^2+(d-7)n+d-2] \ga^{(n+1)}_{AB} + (n-1)s_{AB}^{} \ga^{(n+1)}_{} \\
&- D^C D_C  \ga^{(n+1)}_{AB} + 2D^{}_{(A} D^C \ga^{(n+1)}_{B)C} - D^{}_A D^{}_B \ga^{(n+1)}_{} \ , 
\end{split}
\een
from the $(uA)$ components
\ben\label{einn2}
\begin{split}
0=& 2(d-5-n) D^{}_A \alpha^{(n+1)}_{} - D^C(D^{}_A \beta^{(n+1)}_C - D^{}_C \beta^{(n+1)}_A ) \\
&-n(d-5-n) \beta^{(n+1)}_A  \ , 
\end{split}
\een
from the $(uu)$ component
\ben\label{einn3}
0= D^A D_A \alpha^{(n+1)}_{} - (n+1)(d-4-n) \alpha^{(n+1)}_{} \ , 
\een
from the $(rr)$ component
\ben\label{einn4}
0= -n(n+1) \ga^{(n+1)} \ , 
\een
and from the $(rA)$ components
\ben\label{einn5}
0= [n(d-4)+2(d-3)] \beta^{(n+1)}_{A} + (n+1)D^B \ga^{(n+1)}_{AB} - (n+1) D_A^{} \ga_{}^{(n+1)} \ .
\een
Again we analyze these equations by decomposing the tensors according to lemma~\ref{decomp}. From~\eqref{einn3}, we get $\alpha^{(n+1)}=0$
as long as $n<d-4$ using the spectrum of the Laplacian on scalars, whereas for $n=d-4$, we get $\alpha^{(d-3)} \equiv c$, where $c$ is a constant.
From \eqref{einn2} we then learn that the scalar part, $S$, of $\beta^{(n+1)}_A$ vanishes as long as $d-4 \ge n \neq d-5$. Again from~\eqref{ein2}, 
the vector part $V_A$ 
of $\beta^{(n+1)}_A$ must satisfy
\ben
-D^C D_C V_A = [n^2 - (d-5)n - (d-3)] V_A \ . 
\een
Using the spectrum of the Laplacian on vectors, we see that this equation does not have non-trivial solutions for $n \le d-4$, so the vector part vanishes in this range. 
The tensor part of \eqref{einn1} implies that the tensor part $T_{AB}$ of $\ga^{(n+1)}_{AB}$ satisfies
\ben
-D^CD_C T_{AB} = -[n^2+(d-7)n+d-2] T_{AB} \ . 
\een 
Using the spectrum of the Laplacian on tensors we see that this equation has no non-trivial solutions for $n \le d-4$. 
Therefore, $\ga^{(n+1)}_{AB}$ has no tensor part for $n \le d-4$. We next consider the vector part of~\eqref{einn1}  and use that the vector part 
of $\beta^{(n+1)}_A$ is already known to vanish in this range. The vector part $W_A$ of $\ga^{(n+1)}_{AB}$ is then seen to obey
\ben
[n^2+(d-7)n+d-4] D_{(A}^{} W_{B)} = 0 \ , 
\een
and it follows that $\ga^{(n+1)}_{AB}$ has no vector part in this range either. By~\eqref{einn4}, 
the scalar part(s) $(T, U)$ of $\ga^{(n+1)}_{AB}$ must satisfy $U=0$. 
Inserting this into~\eqref{einn5} then relates $T$ to the scalar part $S$ of $\beta^{(n+1)}_A$ in the following way:
\ben\label{scala}
0= D_A \bigg\{
(d-2)[(d-4)n+2(d-3)]S + (d-3)(n+1)[D^C D_C + d-2]T
\bigg\} \ . 
\een
When $n \neq d-5$, we already know that $S=0$, so it follows that $(D^B D_B  + d-2)T$ is constant, which is possible only when $T$ itself is a constant or an $l=1$
spherical harmonic. But then 
it follows in view of $U=0$  that also the scalar part of $\ga^{(n+1)}_{AB}$ must vanish. For $n=d-5$, we may use another relation between $S$ 
and $T$ from the trace of \eqref{einn1}, namely
\ben
0=D^C D_C \bigg\{
(d-2)(2d-n-6)S + (d-3)[D^A D_A +d-2]T
\bigg\} \ . 
\een
Taking $D^A$ in \eqref{scala} and $n=d-5$, it follows at this order that $(d-2)(d-5) D^C D_C S=0$, hence that $S$ is constant. 
But then it also follows that $D^B D_B[D^A D_A +d-2]T=0$, meaning that $T$ is a linear combination of $l=0$ and $l=1$ spherical harmonics. 
Thus, neither $T$ nor $S$ makes a contribution to the scalar parts of $\beta^{(n+1)}_A$ respectively $\ga^{(n+1)}_{AB}$ also when $n=d-5$. 
 This closes the induction loop.

We  summarize our findings in this section as follows. 

\begin{theorem}\label{thm1}
Let $(\M,g)$ be an asymptotically flat spacetime of even dimension $d \ge 4$ satisfying (i)--(iv), presented in CGN gauge~\eqref{Pen1}, 
with Killing field $K$ that is timelike near $\sI^+$.
Then there exist unique diffeomorphisms $\phi$ and $\psi$ defined near $\sI^+$ which preserve the CGN gauge such that
\begin{enumerate}
\item $\phi$ is generated by an asymptotic boost vector field of type (III),
\item $\psi$ is generated by a linear combination of a supertranslation vector field of type (IIa) and a gauge transformation of type (I),

\item  {\bf In all dimensions $d \ge 4$,} 
the expansion coefficients~\eqref{coeff} of $\phi^* \psi^* g$ satisfy
\ben\label{sd4}
\ga^{(0)}_{AB}=s_{AB}^{} \ , \quad 
\ga^{(d-3)}_{AB}=0 \ , \quad 
\alpha^{(0)} = 1/2 \ , \quad \alpha^{(d-3)} = c \ , \quad 
\beta^{(0)}_A = \beta^{(d-3)}_A = 0 \ , 
\een
where $c$ is a constant. \\
{\bf In $d>4$ and even,} the expansion coefficients~\eqref{coeff} of 
 $\phi^* \psi^* g$ satisfy additionally for all $1 \leq n \leq d-4$,  
\ben\label{sd>}
\ga^{(n)}_{AB} = 0 \ , \quad \beta^{(n)}_A = 0 \ , \quad \alpha^{(n)}_{} = 0\ . 
\een

\item 
$\phi^* \psi^* K=\partial/\partial u$ near $\sI^+$.
\end{enumerate}
\end{theorem}

It follows immediately from this theorem and the definition of the Bondi energy $\EB$ and Bondi linear momentum $\PB$ (see~~\eqref{mux}-\eqref{MBdef} below)
that for the boosted metric $\phi^* \psi^* g$ we have $(\EB, \PB) = (\MB, {\bf 0})$, where 
\ben\label{c}
\MB = \frac{{\rm vol}(S^{d-2})}{4\pi} c \quad (= c \quad \text{in $d=4$}) \ . 
\een
Indeed, we know that $\alpha^{(d-3)}$ for the boosted metric is equal to $c$, 
whereas all other coefficient tensors of the boosted metric appearing in the integrand~\eqref{mux} of the Bondi energy/momentum vanish. 
The Bondi energy/momentum of the original 
metric $g$ are then obtained by reversing the action of the asymptotic boost $\phi$. Hence, if the asymptotic boost has velocity parameter $\v$ 
[see~footnote 10], 
we conclude that the original metric $g$ has Bondi energy resp. momentum 
\ben\label{43}
\EB = \frac{\MB}{\sqrt{1-v^2}} \ , \quad \PB = \frac{\MB\v}{\sqrt{1-v^2}} \ , 
\een 
with $\MB$ related to $c$ as in~\eqref{c}.
Of course, we may also read this as an equation for $\v$ for given $(\EB, \PB)$. 

Next we recall that the boost $\phi$ with velocity parameter $\v$ acts as a conformal transformation 
$\hat \phi$
on the ``celestial sphere'' $S^{d-2}$ (i.e. the cut of $\sI^+$ parameterized by unit vectors $\bf z$) in our conformal Gaussian coordinated system acting by~\eqref{conf}. 
For the CGNCs $(u,r,z^A)$ we have that
\ben
\phi^* r = e^{-\omega} r + O(ur^0) \ , \quad \phi^* u = e^{\omega} u + O(u^2r^{-1}) \ , \quad \phi^* z^A = \hat \phi^* z^A + O(ur^{-1})
\een
and we can use this information to go from $\psi^* \phi^* g$ with coefficients characterized by the previous theorem, back to $g$. 
Exploiting~\eqref{43} to express $\v$ by $(\EB,\PB)$, we immediately get after a short calculation:
\begin{lemma}\label{alphalemma}
Let $(\M, g)$ be a stationary solution to the Einstein vacuum equations near $\sI^+$. Then the expansion coefficient~\eqref{coeff}
$\alpha^{(d-3)}$ for $g$ must be given by
\ben
\alpha^{(d-3)} = c e^{(d-1)\omega} = \frac{4\pi}{{\rm vol}(S^{d-2})}\frac{\MB^d}{(\EB-\PB \cdot {\bf z})^{d-1}}  \ , 
\een 
where $(\EB, \PB)$ is the Bondi energy-momentum of $(\M,g)$, and $\MB = \sqrt{\EB^2-|\PB|^2}$ the Bondi mass. 
\end{lemma}

\section{Asymptotic Expansion of Non-stationary Metrics}\label{s2}

We now contrast the asymptotic expansion of the metric found in stationary case (Thm.~\ref{thm1}) with that in the non-stationary case. As before, we 
assume that $(\M,g)$ is asymptotically flat. To avoid awkward issues coming from the precise behavior of the metric near {\em spatial} infinity, it is convenient to assume, 
additionally, that the metric is stationary in a neighborhood of spatial infinity. Thus we assume in addition to (i)--(iv):
\begin{enumerate}
\item[(v)] There exists a $u_0$ such that $(\M,g)$ is stationary (with Killing field $K$) in a neighborhood of $\sI^+$ for $u$-values $u<u_0$. 
\end{enumerate} 
As already briefly discussed in the Introduction, this new assumption is reasonable because, due to the Corvino-Schoen gluing constructions~\cite{glue1}, \cite{glue2}, one can always glue any portion of initial data to 
an asymptotic end that is exactly equal to a Kerr or Myers-Perry solution near spatial infinity. The evolved metric will then be stationary at $\sI^+$ at early times. Thus, this assumption does not exclude any initial conditions that one might wish to consider in the interior of the spacetime. However, it should be noted that this assumption does exclude the sort of initial conditions that are normally considered in scattering theory where matter comes in from infinity at asymptotically early times\footnote{We will consider such particle scattering initial conditions in the context of linearized gravity in \cite{GHITW}, where it will be shown that the memory effect is absent for $d>4$ in this case as well.}.

For the portion of $\sI^+$ corresponding to $u<u_0$, we can then apply the conclusions of Thm.~\ref{thm1}, because the proof of that theorem was entirely local. 
Therefore, we get an asymptotic boost $\phi$ and an asymptotic supertranslation $\psi$ such that the expansion coefficients of $\psi^* \phi^* g$
described in the theorem vanish up to order $d-3$, and such that $\psi^* \phi^* K = \partial/\partial u$ for $u<u_0$. To the metric $\psi^* \phi^* g$, we can 
therefore further apply Lemma~8 of~\cite{thorne} in order to determine the form of the expansion coefficients in the non-stationary part $u\ge u_0$. 
The conclusions are as follows:

\begin{theorem}\label{thm2}
Let $(\M,g)$ be an asymptotically flat spacetime of even dimension $d \ge 4$ satisfying (i)--(v), presented in CGN gauge~\eqref{Pen1}.
Then there exist unique diffeomorphisms $\phi$ and $\psi$ defined near $\sI^+$ which preserve the CGN gauge such that
\begin{enumerate}
\item $\phi$ is generated by an asymptotic boost vector field of type (III),
\item $\psi$ is generated by a linear combination of a supertranslation vector field of type (IIa) and a gauge transformation of type (I), 
\item {\bf For $d > 4$ and even}, the expansion coefficients~\eqref{coeff} of $\phi^* \psi^* g$ satisfy for any $1 \le n \le (d-4)/2$:
\ben\label{2}
0= \alpha^{(n)}  \ , \quad
0= \beta^{(n)}_A \ , \quad
0= \ga^{(n)}_{AB} \ ,
\een
as well as 
\ben\label{3}
\ga^{(0)}_{AB}=s_{AB}^{} \ , \quad \alpha^{(0)} = 1/2 \ ,  \quad 
\beta^{(0)}_A  = 0 \ . 
\een
Furthermore, for $1 \le n \le d-3$ in the first, and for $1 \le n\le d-4$ in the second equation:
\ben\label{4}
\begin{split}
\beta_A^{(n)} &= -\frac{n}{(n+1)(d-2-n)} D^B\ga_{AB}^{(n)} \quad ,\\
\alpha^{(n)}  &=\frac{n-1}{2n(d-3-n)} D^A\beta_A^{(n)} \quad , 
\end{split}
\een
and for $1 \le n \le d-3$
\ben\label{5}
\ga^{(n)} \equiv s^{AB}_{} \ga^{(n)}_{AB} = 0 \quad .
\een
\item {\bf For $d=4$}, the expansion coefficients of $\phi^* \psi^* g$ satisfy
\ben\label{d4}
\ga^{(0)}_{AB}=s_{AB}^{} \ , \quad \alpha^{(0)} = 1/2 \ , \quad 
\beta^{(0)}_A=0 \ , \quad  \partial_u \ga^{(1)} = 0 \ , \quad \beta^{(1)}_A = -\frac{1}{2} D^B \ga^{(1)}_{AB} \ . 
\een
\end{enumerate}
\end{theorem}

Using this result, one can determine how the conserved quantities associated with the asymptotic symmetries 
(II)--(IV) are related to the expansion coefficients~\eqref{coeff}. Here we only recall the situation for ``ordinary'' infinitesimal 
translations  parameterized by a vector field $X$ as in  cases (IIb,c), respectively. Define the ``mass aspect'' by
\ben\label{mux}
\mu = \ (d-2) \bigg[ \alpha^{(d-3)}  - \frac{1}{8(d-3)} \ga^{(d/2-1)AB}_{} \partial_u \ga^{(d/2-1)}_{AB} \bigg]  \ . 
\een 
Then the Hamiltonian (generator) of an infinitesimal translational symmetry in class (IIb,c) is given by~\cite{thorne}
\ben\label{Mdef}
{\mathcal H}_X(u) =  \frac{1}{8\pi} \int_{S^{d-2}(u)} T \cdot \mu  \sqrt{s} \d^{d-2} z \ . 
\een
where $S^{d-2}(u)$ is a cut of $\sI^+$ at the value $u$ of the affine parameter, and where
 $T=1$ for time translations (IIc), whereas $T(z) = {\bf a}\cdot {\bf z}$ for spatial translations in the direction $\bf a$ (IIb). 

This formula for the Hamiltonian generator holds for all even $d \ge 4$. For $d=4$, there is also a generator conjugate to 
supertranslations (IIa). It is given by simply taking $T$ to be a function on $S^2$ orthogonal to the $l=0,1$ spherical harmonics, see class (IIa). 
In even $d>4$ there is {\em no} Hamiltonian generator conjugate to the BMS supertranslations, as follows from the analysis of~\cite{HI}. This is closely related to the 
fact that these generators are not to be viewed as symmetries on the phase space of general relativity in $d>4$, as we will discuss further at the end of section \ref{summary} below. 

The generators have the following interpretation:
\ben\label{MBdef}
{\mathcal H}_X(u) =
\begin{cases}
\EB(u) & \text{Bondi energy for $X$ in case (IIc),}\\
{\bf a} \cdot \PB(u) & \text{Bondi linear momentum for $X$ in case (IIb), ($T(\z)=\z \cdot {\bf a}$).} 
\end{cases}
\een
For the stationary case, this interpretation can be confirmed by an explicit computation
using lemma~\ref{alphalemma} and thm.~\ref{thm1} in the expression~\eqref{mux}. 
The ``flux'' formula is, in all\footnote{In $d=4$, there is also a flux formula for supertranslation charges, case (IIa). 
The flux is now $$-\frac{1}{32\pi} \int_{u_0}^{u_1} \left( \int_{S^{2}(u)} N^{AB} (N_{AB} + 2D_AD_B)T \ \sqrt{s} \d^{2} z \right) \d u \,.$$
The formula follows from the identity given below eq.~\eqref{below}.}  cases (IIb,c)
\ben\label{fluxlaw}
{\mathcal H}_X \ \Bigg|_{u_0}^{u_1} = -\frac{1}{32\pi} \int_{u_0}^{u_1} \left( \int_{S^{d-2}(u)} T  N^{AB} N_{AB} \ \sqrt{s} \d^{d-2} z \right) \d u \ , 
\een
where the news tensor is
\ben
N_{AB} = -\partial_u \ga^{(d/2-1)}_{AB} \ . 
\een
An invariant formula for ${\mathcal H}_X$ in $d=4$ was given first by Geroch~\cite{G}, and the 
relationship to the Hamiltonian framework was later clarified in~\cite{WZ} and~\cite{Ashtekar}, see also~\cite{Dray}. In higher dimensions, 
an invariant formula for ${\mathcal H}_X$ was derived in~\cite{HI} (using the framework of~\cite{WZ}). That the above expression~\eqref{MBdef} for $\mu$ is equivalent to the 
invariant formulas can be seen using Thm.~\ref{thm2}, see~\cite{thorne} for details\footnote{In that reference, only the case 
$T=const.$, i.e. case (IIc) was treated explicitly. But the result easily generalize to the other cases, including to supertranslations in $d=4$}. 
 
 We also note that we always have the positive energy theorem $\EB \ge |\PB|$, see~\cite{horowitz, yau, ludvigsen,chrusciel} in $d=4$ and~\cite{thorne} in $d>4$. 
 In $d>4$ the 
constructions are more complicated since e.g. $\alpha^{(d-3)}$ is {\em not} the leading order expansion coefficient. 
 
\section{BMS Symmetry and Memory}\label{memory}

\subsection{Relation between metrics before and after a radiation burst}

Comparing the stationary situation  described by Thm.~\ref{thm1} and the non-stationary (``radiating'') situation
described by Thm.~\ref{thm2}, we can now explain in detail the crucial qualitative difference between $d=4$ and higher $d>4$. This difference is perhaps appreciated 
best if we consider a solution which is stationary near $\sI^+$ for early times $u<u_0$ as we have
been assuming, and becomes stationary again at late times\footnote{The assumption that the spacetime becomes exactly stationary again at a finite time $u_1$ is, of course, an idealization, i.e., one should consider the limit $u_1 \to \infty$.} $u>u_1$. By Thm.~\ref{thm2}, the order at which 
gravitational radiation occurs, i.e. where the expansion coefficients~\eqref{coeff} can be time-dependent, is $n=d/2-1$. 

Now, by Thm.~\ref{thm1}, we can find an 
asymptotic boost and an asymptotic supertranslation, denoted collectively $f_0$, such that the expansion coefficients~\eqref{coeff} of $f^*_0 g$  
vanish up to order $n \le d/2-1$ for  $u<u_0$, i.e. {\em before} the ``radiation epoch''. Similarly, {\em after} the radiation epoch, i.e. for $u>u_1$, we can again find an 
asymptotic boost and an asymptotic supertranslation, denoted collectively $f_1$, such that the expansion coefficients of $f^*_1 g$ vanish up to order $n \le d/2-1$. 
So far, there is no difference between $d=4$ and $d>4$. However, we now claim:

\begin{lemma}\label{thm3}
\begin{enumerate}
\item {\bf In even $d>4$,} it is true that $f_0=f_1$, so the expansion coefficients~\eqref{coeff} of $g$ for $u< u_0$ agree with those for $u>u_1$ up to any
$n \le d/2-1$. In particular, we can apply a {\em single} supertranslation boost $f$ such that $f^* g$ has 
vanishing non-trivial expansion coefficients~\eqref{coeff} up to order $n \le d/2-1$ for $u>u_1$ {\em and} $u<u_0$.

\item {\bf In $d=4$,} we have $f_0 \neq f_1$ in general, so, in general, the expansion coefficients~\eqref{coeff} of $g$ at order $n=d/2-1=1$ for 
$u<u_0$ and $u>u_1$ disagree. $f=f_0 \circ f_1^{-1}$ is the product $f=\phi \circ \psi$ of an asymptotic boost $\phi$ (III), and a combination $\psi$ of an
asymptotic BMS supertranslation (IIa) and a gauge transformation (I) such that 
the expansion coefficients of $g$ before $u_0$ agree with those of $f^*g$ after $u_1$ up to order $n=1$ (up to a change in $\alpha^{(1)}$ as in lemma~\ref{alphalemma}). 

\item
{\bf In $d=4$,} let $T: S^2 \to {\mathbb R}$ be  the parameter of the BMS supertranslation $\psi$ relating the metrics before and after the radiation epochs in 2). 
Then we have for the expansion coefficients of $g$:
\ben
\begin{split}\label{difference}
\ga^{(1)}_{AB}  \Bigg|_{u<u_0}^{u>u_1}   &=  2 \left( D_A D_B T - \frac{1}{2} s_{AB} D^C D_C T \right) \ , \\
\beta^{(1)}_{A} \Bigg|_{u<u_0}^{u>u_1}  &=  -\frac{1}{2} D_A\left( D^B D_B T + 2T \right) \ , \\
\alpha^{(1)}_{}  \Bigg|_{u<u_0}^{u>u_1} &= \frac{1}{4} D^A D_A(D^B D_B + 2)T + 4 \pi \F
\end{split}
\een
where 
$\F \le 0$ is the flux of gravitational radiation per angle $z \in S^2$, defined by 
\ben\label{fperangle}
\F(z) = -\frac{1}{32 \pi} \int_{u_0}^{u_1} N_{AB} N^{AB}(u,z) \ \d u \ , 
\een
in terms of the Bondi news tensor $N_{AB}$. 
\end{enumerate}
\end{lemma}
{\em Proof:}  1) For stationary metrics, having applied $f_0$ resp. $f_1$, we learn from Thm.~\ref{thm1} that all non-trivial 
expansion coefficients of $f^*_0 g$ resp. $f^*_1 g$ vanish up to order $n<d-3$ and $u<u_0$ resp. $u>u_1$. This gives, for instance
\ben
\begin{split}\label{difference1}
\ga^{(1)}_{AB} \Bigg|_{u<u_0}   &=  -2(D_A D_B + s_{AB}) T_0 + 2s_{AB} S_0 \ , \\
\ga^{(1)}_{AB} \Bigg|_{u>u_1}   &=  -2(D_A D_B + s_{AB}) T_1 + 2s_{AB} S_1 \ , 
\end{split}
\een
for the expansion coefficients of $g$, where $T_i, S_i: S^{d-2} \to {\mathbb R}$ are the parameters of the transformations $f_i$ generated by the vector fields of type (I) or (II).
However, we also learn from Thm.~\ref{thm2} that all non-trivial 
expansion coefficients of $f^*_0 g$ and $f^*_1 g$ vanish up to order $n<d/2-1$ for all $u$. Since for $d>4$ it is obviously true that
 $d/2-1>1$, it follows in those dimensions that $T_1=T_0$ and $S_1=S_0$. \\
 2) On the other hand, in $d=4$, we have $d/2-1=d-3=1$, so 
 this conclusion cannot be drawn. Instead in general we only learn that $f=f_0 \circ f_1^{-1}$ is the product of the transformations claimed in 2). The rest of the statements 
 follows again from Thm.~\ref{thm1} and lemma~\ref{alphalemma}. \\
 3) Even in $d=4$, we still know that the trace $\ga^{(1)}$ is independent of $u$ by~\eqref{d4}, so  
 the difference~\eqref{difference} has vanishing trace, which allows us to determine $S=S_0-S_1$ in terms of $T=T_0-T_1$ giving $S=\half(D_A D^A + 2)T$.
 The first relation in ~\eqref{difference} then follows from Thm.~\ref{thm1} and the fact that a BMS supertranslation generated by $X$ as in (IIa) 
 shifts $\ga^{(1)}_{AB}$ precisely by $-2(D_A D_B + s_{AB})T$. 
 The second relation in~\eqref{difference} follows from the first one e.g. from the last relation in~\eqref{d4}. 
To prove the last relation in~\eqref{difference}, we first consider at order $r^{-2}$ the $(rr)$ component of Einstein's equation 
\ben
\ga^{(2)} = -\frac{1}{4} \ga_{AB}^{(1)} \ga^{(1)AB}_{} \ , 
\een
as well as the $(uu)$ component
\ben
\begin{split}
0 =& -\partial_u^2 \ga^{(2)}_{} - \partial_u^2 \ga_{AB}^{(1)} \ga^{(1)AB}_{} - \frac{1}{2} \partial_u \ga_{AB}^{(1)} \partial_u \ga^{(1)AB}_{}\\
&+4 \partial_u \alpha^{(1)}_{}
 + 2 D^A \partial_u \beta^{(1)}_A \ . 
 \end{split}
 \een
Eliminating $\ga^{(2)}$ gives
\ben\label{below-1}
0 = - \frac{1}{2} \partial_u^2 \ga_{AB}^{(1)} \ga^{(1)AB}_{} +4 \partial_u \alpha^{(1)}_{}
 + 2 D^A \partial_u \beta^{(1)}_A \ . 
\een
Substituting the last equation in~\eqref{d4} to eliminate $D^A \beta^{(1)}_A$ next gives
\ben\label{below}
D^A D^B \partial_u \ga^{(1)}_{AB} = \partial_u \left( 4 \alpha^{(1)}_{} - \frac{1}{2} \partial_u \ga_{AB}^{(1)} \ga^{(1)AB}_{} \right) + \frac{1}{2} \partial_u \ga_{AB}^{(1)} \partial_u \ga^{(1)AB}_{} \ . 
\een
[Note that this equation can also be written as $\partial_u \mu = -\frac{1}{4} (N_{AB} + 2 D_A D_B) N^{AB}$, which is equivalent to the mass loss formula in $d=4$.] 
Now we integrate this equation from $u_0$ to $u_1$ and use the first equation in~\eqref{difference} as well as $\partial_u \ga^{(1)}=0$, to give
\ben\label{56}
\alpha^{(1)}_{AB}  \Bigg|_{u<u_0}^{u>u_1} = \frac{1}{4} D^A D_A(D^B D_B + 2)T - \frac{1}{8} \int_{u_0}^{u_1} \partial_u \ga_{AB}^{(1)} \partial_u \ga^{(1)AB}_{}  \ du \ . 
\een
This immediately gives the last equation in~\eqref{difference} after substituting the definition of the Bondi news tensor, $N_{AB} = - \partial_u \ga_{AB}^{(1)}$ and of $\F$. 
\qed

The next proposition gives the relation between the BMS supertranslation parameter $T$ relating the ``BMS-frames'' before and after the radiation, the Bondi 4-momentum 
before and after the radiation, and the total flux per angle $\F(\z)$ defined by \eqref{fperangle}. Using the standard relationship between  $l=0,1,\dots$ 
spherical harmonics on $S^2$, degree $l$ homogeneous harmonic polynomials on $\mathbb{R}^3$, and rank $l$ totally symmetric, trace-free tensors on $\mathbb{R}^3$, we 
can write
\ben\label{multipole}
\F(\z) = \sum_{l=0}^\infty \F^{(l)}_{i_1 \dots i_l} \z^{i_1} \cdots \z^{i_l} \ , 
\een
where each $\F^{(l)}_{i_1 \dots i_l} $ is a totally symmetric, trace-free rank $l$ tensor on $\mathbb{R}^3$. We substitute this relation into 
the last item of lemma~\ref{thm3}, we use lemma~\ref{alphalemma} to substitute $\alpha^{(1)}$, we use the formula
\ben
\half D^C D_C (D^B D_B + 2) \bigg( (\EB - \PB \cdot \z) \log (\EB - \PB \cdot \z) \bigg) 
= \frac{\MB^4}{(\EB - \PB \cdot \z)^3} - \EB - 3\PB \ ,  
\een
and we note that $4\pi \F(\z) - (\EB + 3\PB \cdot \z) |_{u_0}^{u_1}$ has no $l=0,1$ spherical harmonic contribution. (The last claim can be proven by applying the mass loss formula~\eqref{fluxlaw} to an arbitrary combination $T$ of $l=0,1$ spherical harmonics.) If we finally note the identity $D^A D_A(D^C D_C +2) = (L-1)L(L+1)(L+2)$, where $L$ is 
the total angular momentum operator on $S^2$, then we get:
\begin{proposition}
In $d=4$, the BMS supertranslation parameter $T:S^2 \to {\mathbb R}$  relating the metrics before and after the radiation epochs is given by 
\ben
\begin{split}
T(\z) =
& 2(\EB - \PB \cdot \z) \log (\EB - \PB \cdot \z) \bigg|_{u_0}^{u_1}\\
& - 16\pi \sum_{l=2}^\infty \frac{1}{ \prod_{s=1}^4 (l-2+s)}  \F^{(l)}_{i_1 \dots i_l} \z^{i_1} \cdots \z^{i_l} \ , 
\end{split}
\een
where $\F(\z)$ is the flux per angle $\z \in S^2$, see~\eqref{fperangle}, $\F^{(l)}_{i_1...i_l}$ are its $l$-th multi-pole moments~\eqref{multipole}
and $(\EB, \PB)(u_i)$ is the Bondi 4-momentum before and after the radiation epoch. 
\end{proposition}

\medskip

\noindent
{\bf Remark:} The supertranslation $T$ encodes ``information'' about the flux of gravitational radiation emitted between the stationary eras $u < u_0$ and $u > u_1$. However, this information is not locally measurable by observers with access only to the late time stationary era $u > u_1$, i.e., to determine $T$ one would need access to the preferred stationary frames of theorem \ref{thm1} in both the early time era $u < u_0$ and the late time era $u > u_1$.
 
\subsection{Absence of a Memory Effect in $d>4$}

Our conclusions expressed in lemma~\ref{thm3} can be stated more physically and geometrically by saying that: 1) in $d>4$ there is no ``memory effect,'' and 2) in $d=4$ the gravitational memory of a distribution of test masses can be characterized by a BMS supertranslation.  To study the influence of a time-dependent gravitational field on test-masses, we study as usual the geodesic deviation 
equation
\ben
(\tau^a \nabla_a)^2 \xi^b =-R_{acd}{}^b \tau^a \tau^d \xi^c \ , 
\een
where $\tau$ is a vector field that is tangent to a congruence of time like geodesics, and $\xi$ is the deviation vector field, see e.g. sec.~3.4 of \cite{wald}. We are interested in a congruence in the asymptotic region, which is, to leading order, given by an asymptotic time-like translation, i.e. 
a time-like linear combination of vector fields of type (IIb,c). 
For simplicity, we first consider the case (IIc), so we can say that $\tau = \partial/\partial u + O(r^{-2})$. 
The relevant Riemann tensor components are then those with two $u$ indices, which using thm.~\ref{thm2}, are found to be
\ben
\begin{split}
R_{uAuB} &= -\frac{1}{2} r^{3-d/2} \ \partial_u^2 \ga^{(d/2-1)}_{AB} + O(r^{2-d/2}) \ , \\
R_{uruA} &=  O(r^{1-d/2}) \ , \\
R_{urur} &= O(r^{-1-d/2}) \ , 
\end{split}
\een
near $\sI^+$. 
We can use these relations to integrate the geodesic deviation equation twice. This process basically removes the two derivatives in the first equation, 
whereas the second and third equations are telling us that the other Riemann components are sub leading. Altogether, we find:
\ben
\xi \ \Bigg|_{u<u_0}^{u>u_1} = r^{-d/2+1} \Delta(\xi) \ \Bigg|_{u<u_0} + O(r^{-d/2}) \ , 
\een
where the linear map $\Delta$ is the ``memory-'' or ``displacement tensor'', given by\footnote{Note that $\Delta$ has a well-defined restriction to $\sI^+$.}
\ben
\Delta \equiv -\frac{1}{2} s^{AC} \ga^{(d/2-1)}_{CB} \, \frac{\partial}{\partial x^A} \otimes \d x^B  \Bigg|_{u<u_0}^{u>u_1} \ .
\een
By lemma~\ref{thm3}, the right side is given by $-1/2$ times the first equation~\eqref{difference} in $d=4$
and by zero in $d>4$. We thus arrive at the following conclusion:

\begin{proposition}\label{propmem}
The displacement tensor for a congruence of time-like geodesics with tangent $\tau = \partial/\partial u + \dots$ 
near $\sI^+$ is given by $\Delta = \Delta^A{}_B \partial_A \otimes \d x^B$, where
\ben\label{displ}
\Delta_{AB} = 
\begin{cases}
- (D_A D_B T - \frac{1}{2} s_{AB} D^C D_C T) & \text{for $d=4$,}\\
0 & \text{for $d>4$.}
\end{cases}
\een
where $T:S^2 \to {\mathbb R}$ is the parameter of the asymptotic BMS supertranslation relating the metrics before and after the radiation epoch. 
\end{proposition}

In case we have a general time-like congruence, denoted $\hat \tau$, in the asymptotic region can be dealt with essentially by applying 
an asymptotic boost~[see~footnote 10] 
to the previous steps.  More precisely, let $\phi$ by an asymptotic boost with velocity parameter $\v$ chosen so that 
$\hat \tau = \phi^* \tau$, where $\tau = \partial/\partial u + \dots$. Due to general covariance, the displacement tensor calculated for $\hat \tau$ and metric $g$ is equal to the pull back via $\phi^*$ of the displacement tensor $\hat \Delta$ calculated for $\tau$ and $\hat g = \phi^{-1 *} g$. On the 
other hand, $\hat \Delta$ is given by the same formula as in the previous proposition, except that $T$ is replaced by the corresponding $\hat T$ for $\hat g$, which 
in turn is given by $\hat T = e^{-\omega_{\phi^{-1}}} T \circ \hat \phi^{-1}$, where $\hat \phi$ is the conformal transformation of $S^2$ induced by $\phi$, and $e^{\omega_\phi}$ the corresponding conformal factor, see~\eqref{conf}.  Applying $\phi^*$, and using $\omega_{\phi^{-1}} = -\omega_\phi \circ \hat \phi^{-1}$, we learn that 
the displacement tensor $\Delta_{AB}$ for $\hat \tau$ and $g$ is given by the same formula as~\eqref{displ}, except that $T$ is replaced by 
\ben
T(\z) \to \frac{ T(\z)\sqrt{1-v^2}}{1-\v \cdot \z} \ ,  
\een
where $\v$ is the speed of $\hat \tau$ in the spatial direction. 

\section{Summary}\label{summary}

Let us close by summarizing the reasons why there are no BMS supertranslations in $d>4$. 

\begin{enumerate}
\item
In an asymptotically flat spacetime $(\M,g)$ of even dimension $d\ge 4$ that is stationary near spatial infinity (i.e. for $u<u_0$), we can find 
an asymptotic boost and an asymptotic BMS supertranslation such that the metric has vanishing non-trivial expansion coefficients~\eqref{coeff} up to order $n \le d/2-1$ 
for $u<u_0$. In $d>4$, the same statement is then automatically true for any later stationary epoch. For $d=4$, this statement is false and the order $n=d/2-1=1$ 
coefficients are non-zero. They are however pure gauge and in fact correspond to a ``pure BMS supertranslation'', parameterized by a non-trivial function $T$ on $S^2$. 

\item 
This function $T$ is a ``potential'' of the displacement tensor in the memory effect by Prop.~\ref{propmem} in $d=4$, so in this sense we could in principle 
``measure'' the BMS supertranslation by a comparison of the relative displacement of test particles at early and late times. In $d>4$, the displacement tensor is zero, i.e. there is no gravitational memory at the order where 
tidal effects due to gravitational radiation become visible. (Of course, geodesics would 
be subject to relative displacements resulting from ordinary tidal effects. However, such tidal effects occur at subleading $1/r$-order compared to the ones caused by gravitation radiation.) 

\item
In dimension $d>4$, we  can thus impose as a reasonable gauge condition that, for spacetimes which are stationary near spatial infinity (i.e. for $u<u_0$), 
the metric has vanishing non-trivial expansion coefficients~\eqref{coeff} up to order $n \le d/2-1$ 
for $u<u_0$. The remaining asymptotic symmetries then correspond to ordinary asymptotic Poincare symmetries. By item 1), this is not natural in $d=4$. 

\item
In $d=4$, we can find Hamiltonian generators conjugate to asymptotic BMS symmetries. By the analysis of~\cite{HI,WZ}, the same is not possible in $d>4$.  Indeed, it was argued that 
in those papers that, in order to define a charge conjugate to an infinitesimal gauge transformation, there should at the very least be a finite flux of symplectic current at $\sI^+$, where the symplectic current is defined by~\cite{WZ}
\ben
j^a = g^{abcdef}(\delta_1 g_{bc} \nabla_d \delta_2 g_{ef} - (1 \leftrightarrow 2)) 
\een
for a pair of two solutions of the linearized Einstein equations. By assumption, they should satisfy the linearized asymptotic flatness conditions, and 
then thereby also the linearized version of Thm.~\ref{thm2}. However, one easily sees that the symplectic flux across $\sI^+$ is then infinite in $d>4$, unless we assume that we are in the special gauge characterized by eq.~\eqref{2} of that theorem. But imposing this gauge by construction precisely removes the ``supertranslation gauge mode''
so we are no longer able to define generators conjugate to asymptotic BMS supertranslations $X$ (IIa).

\end{enumerate}

\noindent
{\bf Acknowledgements:} 
S.H. thanks P. Chrusciel for explanations about the gluing theorems in higher dimensions.
R.M.W. wishes to thank Leipzig University for hospitality. The research of R.M.W. was supported by NSF grant PHY 15-05124 to the University of Chicago. 
The work of A.I. was supported in part by JSPS KAKENHI Grants 
No.~15K05092 and No.~26400280.

\end{document}